\documentclass[final,3p,times,onecolumn,sort&compress,hidelinks]{elsarticle}
\usepackage[utf8]{inputenc}
\usepackage{bm}
\usepackage[dvipsnames]{xcolor}
\usepackage{multicol}
\usepackage{multirow}
\usepackage[normalem]{ulem}
\usepackage{subfig}
\usepackage{graphicx}
\usepackage{amsmath}

\def\kT{k_T}
\def\pperp{p_T}

\def\avkT{\la \kT^2 \ra}
\def\avpperp{\la \pperp^2 \ra}

\newcommand{\la}{\langle}
\newcommand{\ra}{\rangle}

\begin{document}

\rightline{JLAB-THY-21-3310}

\title{Electron-Ion Collider impact study on the tensor charge of the nucleon}
\author[label1]{Leonard Gamberg}\ead{lpg10@psu.edu}
\author[label2,label3,label4]{Zhong-Bo Kang}\ead{zkang@g.ucla.edu}
\author[label5]{Daniel Pitonyak}\ead{pitonyak@lvc.edu}
\author[label1,label6]{Alexei Prokudin} \ead{prokudin@jlab.org}
\author[label6]{Nobuo Sato} \ead{nsato@jlab.org}
\author[label7]{Ralf Seidl} \ead{rseidl@ribf.riken.jp}

\address[label1]{Division of Science, Penn State University Berks, Reading, Pennsylvania 19610, USA}
\address[label2]{Department of Physics and Astronomy, University of California, Los Angeles, California 90095, USA}
\address[label3]{Mani L.~Bhaumik Institute for Theoretical Physics, University of California, Los Angeles, California 90095, USA}
\address[label4]{Center for Frontiers in Nuclear Science, Stony Brook University, Stony Brook, New York 11794, USA}
\address[label5]{Department of Physics, Lebanon Valley College, Annville, PA 17003, USA}
\address[label6]{Theory Center, Jefferson Lab, 12000 Jefferson Avenue, Newport News, Virginia 23606, USA}
\address[label7]{RIKEN BNL Research Center, Upton, New York 11973, USA}

\begin{abstract}
\noindent 
In this letter we study the impact of the Electron-Ion Collider (EIC) on the phenomenological extraction of the tensor charge from a QCD global analysis of single transverse-spin asymmetries (SSAs).  We generate EIC pseudo-data for the Collins effect in semi-inclusive deep-inelastic scattering for proton and $^{3\!}He$ beams  across multiple center-of-mass energies.  We find a significant reduction in the uncertainties for the up, down, and isovector tensor charges that will make their extraction from EIC data on SSAs as precise as current lattice QCD calculations. We also analyze the constraints placed by future data from the proposed SoLID experiment at Jefferson Lab, discuss its important complementary role to the EIC, and present the combined impact from both facilities.
\end{abstract}

\maketitle

\section{Introduction}
The tensor charge $g_T$ is one of the fundamental charges of the nucleon~\cite{Ralston:1979ys,Jaffe:1991kp,Jaffe:1991ra,Cortes:1991ja,Gamberg:2001qc} and, arguably, the least known.   
The uniqueness of $g_T$ is also that it sits at the intersection of three key areas of nuclear physics:~3-dimensional tomography of the nucleon~(see, e.g.,~\cite{Anselmino:2013vqa,Goldstein:2014aja,Radici:2015mwa,Kang:2015msa,Radici:2018iag,Benel:2019mcq,DAlesio:2020vtw,Cammarota:2020qcw}), searches for beyond the Standard Model (BSM) physics~(see, e.g.,~\cite{Courtoy:2015haa,Yamanaka:2017mef,Gao:2017ade,Gonzalez-Alonso:2018omy}), and {\it ab initio} approaches like lattice QCD or Dyson-Schwinger Equations~(see, e.g.,~\cite{Gupta:2018qil,Yamanaka:2018uud,Hasan:2019noy,Alexandrou:2019brg,Pitschmann:2014jxa}).  The focus of this letter will be on accessing the tensor charge  through the first avenue, namely, an analysis of single transverse-spin asymmetries (SSAs) that are sensitive to the 3-dimensional structure of the nucleon.  In particular, one can compute $g_T$ from an integral of the transversity parton distribution function (PDF) $h_1(x)$~\cite{Ralston:1979ys,Jaffe:1991kp,Jaffe:1991ra,Cortes:1991ja,Gamberg:2001qc} over the parton momentum fraction $x$:
\begin{equation}
    g_T = \delta u - \delta d\quad {\rm where} \quad \delta u = \int_0^1 \! dx\,(h_1^u(x)-h_1^{\bar{u}}(x))\,, \quad \delta d = \int_0^1 \! dx\,(h_1^d(x)-h_1^{\bar{d}}(x))\,, \label{e:gT}
\end{equation}
while $u$ and $d$ represent up and down quarks, respectively.

Phenomenological extractions of the tensor charge have typically fallen into two main categories.  The first are those studies that use transverse momentum dependent (TMD) observables like the Collins effect in semi-inclusive deep-inelastic scattering (SIDIS)~\cite{Airapetian:2004tw, Alekseev:2008aa,Airapetian:2010ds,Qian:2011py,Adolph:2014zba,Zhao:2014qvx,Airapetian:2020zzo} and semi-inclusive electron-positron annihilation to almost back-to-back hadrons (SIA)~\cite{Seidl:2008xc,TheBABAR:2013yha,Aubert:2015hha,Ablikim:2015pta,Li:2019iyt}, which allow for the transversity TMD PDF $h_1(x,k_T^2)$ and Collins TMD fragmentation function~(FF) $H_1^\perp(z,p_T^2)$ (defined below) to be fit simultaneously~\cite{Anselmino:2013vqa,Kang:2015msa,DAlesio:2020vtw,Cammarota:2020qcw}. We note that the Collins effect for hadron-in-jet measurements from proton-proton collisions is also sensitive to the coupling of $h_1(x,k_T^2)$ and $H_1^\perp(z,p_T^2)$~\cite{Adamczyk:2017wld,Kang:2017btw,DAlesio:2017bvu,Kang:2020xyq}.  In addition, one can use Generalized Parton Distributions (GPDs) to extract the tensor charge~\cite{Goldstein:2014aja}. The second category is those analyses that use di-hadron observables in SIDIS~\cite{Airapetian:2008sk,Adolph:2012nw,Adolph:2014fjw,Braun:2015baa}, SIA~\cite{Vossen:2011fk}, and proton-proton collisions~\cite{Adamczyk:2015hri,Adamczyk:2017ynk}, where the collinear transversity PDF $h_1(x)$ and di-hadron FF $H_1^\sphericalangle(z,M_h)$ can be fit simultaneously~\cite{Radici:2015mwa,Radici:2018iag,Benel:2019mcq}.  Generally the extraction of the tensor charge from both the TMD (SIDIS+SIA) and di-hadron approaches~\cite{Anselmino:2013vqa,Kang:2015msa,DAlesio:2020vtw,Radici:2015mwa,Radici:2018iag,Benel:2019mcq} have shown tension with lattice QCD calculations at the physical point~\cite{Gupta:2018qil,Hasan:2019noy,Alexandrou:2019brg}.  However, we note the study in Ref.~\cite{Lin:2017stx} found that there do exist solutions for $h_1(x,k_T^2)$ and $H_1^\perp(z,p_T^2)$ that can successfully describe both Collins effect SIDIS measurements and lattice data. 

Moreover, there was a recent global analysis of SSAs performed in Ref.~\cite{Cammarota:2020qcw} (JAM20), which included not only Collins effect SIDIS and SIA data but also Sivers effect SIDIS and Drell-Yan measurements~\cite{Airapetian:2009ae, Alekseev:2008aa,Qian:2011py,Adolph:2014zba,Zhao:2014qvx,Adamczyk:2015gyk,Adolph:2016dvl,Aghasyan:2017jop,Airapetian:2020zzo} as well as proton-proton $A_N$ data~\cite{Lee:2007zzh,Adams:2003fx,Abelev:2008af,Adamczyk:2012xd}.  The JAM20 results found for the first time an agreement between experimental data and lattice QCD (without including lattice data in the fit) for $\delta u$, $\delta d$, and $g_T$, as calculated in Eq.~(\ref{e:gT}).  The crucial aspect that allowed for such an agreement was the inclusion of $A_N$ data.  This observable is {\it collinear} twist-3~\cite{Qiu:1998ia, Kouvaris:2006zy, Koike:2009ge, Kang:2011hk, Metz:2012ct, Beppu:2013uda} and dominated by a term that couples $h_1(x)$ to the quark-gluon-quark FFs $H_1^{\perp(1)}(z)$ and $\tilde{H}(z)$~\cite{Kanazawa:2014dca,Gamberg:2017gle,Cammarota:2020qcw}.  The function $H_1^{\perp(1)}(z)$ is the first moment of the Collins TMD FF, and $\tilde{H}(z)$ generates the $P_{hT}$-integrated SIDIS
$A_{UT}^{\sin\phi_S}$ asymmetry, where $P_{hT}$ is the transverse momentum of the hadron w.r.t.~the momentum of the virtual photon, by again coupling with $h_1(x)$~\cite{Bacchetta:2006tn}.

Furthermore, the JAM20 results, due to the inclusion of $A_N$ data, also give the most precise phenomenological extraction of $g_T$ to date:~$g_T=0.87(11)$.  Nevertheless, the error in $g_T$, along with those for $\delta u, \delta d$ (JAM20 values are $\delta u = 0.72(19), \delta d = -0.15(16)$) are still much larger ($\sim \!\!12\% $ for $g_T$, $\sim \!\! 25\%$ for $\delta u$, and $\sim \!\!100\%$ for $\delta d$) than the uncertainties from lattice QCD calculations ($\lesssim \!5\% $ for all of $\delta u, \delta d$, and $g_T$)~\cite{Gupta:2018qil,Hasan:2019noy,Alexandrou:2019brg}.  The main cause of the uncertainty for phenomenological computations is that they rely on integrals of $h_1(x)$ over the entire $x$ region from 0 to 1 (see Eq.~(1)).  However, current SIDIS measurements only cover a region $0.02 \lesssim x\lesssim 0.3$.  This leaves the transversity PDF basically unconstrained in the small-$x$ and large-$x$ regimes.  The inclusion of  $A_N$ data does give some further constraints in the larger-$x$ region since in that observable one integrates from $x_{min}$ to 1, where $0.2\lesssim x_{min}\lesssim 0.7$.  We also mention that a first principles calculation of the small-$x$ asymptotics of the valence transversity TMD PDF has been performed in Ref.~\cite{Kovchegov:2018zeq}.  Nevertheless, one clearly needs very precise data at both $x\lesssim 0.02$ and $x\gtrsim 0.3$ in order to significantly reduce the uncertainties in phenomenological extractions of the tensor charge.

The future Electron-Ion Collider (EIC)~\cite{Boer:2011fh,Accardi:2012qut} at Brookhaven National Laboratory will make the most precise SIDIS measurements at small $x$ (down to $x\!\sim\! 10^{-4}$) while also increasing the precision of the data in the region up to $x\sim 0.5$.  The 12 GeV program currently underway at Jefferson Lab (JLab)~\cite{Dudek:2012vr} will make precision measurements up to $x\sim 0.6$ and smaller values of $Q^2$.  In terms of the tensor charge, future data from the proposed SoLID experiment at JLab~\cite{Chen:2014psa,Ye:2016prn} will offer  substantial constraints in this region. Therefore, we separately assess the impact of SIDIS Collins effect pseudo-data from SoLID (for the ``enhanced'' scenario)~\cite{Gao:privcom}.  We also note that precision measurements from Belle-II on the Collins effect in SIA will affect extractions of the tensor charge due to reducing the uncertainties in the Collins TMD FF~\cite{Kou:2018nap}.  
The goal of this letter is to perform an impact study of future EIC data on the tensor charge of the nucleon using the JAM20 results as a baseline.  In Sec.~\ref{s:EICdata}, we discuss the EIC pseudo-data used in the analysis.  This includes both proton and $^{3\!}He$ beams across multiple center-of-mass (CM) energies.   In Sec.~\ref{s:pheno}, we include these pseudo-data in the global analysis of Ref.~\cite{Cammarota:2020qcw} and, from the newly extracted transversity PDF, compute the tensor charges $\delta u, \delta d$, and $g_T$ and compare them to those of recent lattice QCD calculations. In Sec.~\ref{s:SoLID}, since the proposed SoLID experiment itself would give significant impact on the tensor charge at large $x$, we perform a similar analysis on its pseudo-data~\cite{Gao:privcom}. We also discuss the important complementary role of SoLID to the EIC if one is to obtain an accurate and, as much as possible, unbiased phenomenological extraction of the tensor charge. Finally, we summarize our results and discuss the future outlook in Sec.~\ref{s:concl}.  

\section{Generating EIC Pseudo-Data \label{s:EICdata}}
The EIC will provide data sensitive to the transversity PDF through  SSAs in single-hadron and di-hadron reactions.  For the former, measurements will be made of the Collins effect $A_{UT}^{\sin(\phi_h+\phi_S)}$ in $e+N^\uparrow \to e+h+X$, where $\phi_h$ ($\phi_S$) is the azimuthal angle of the outgoing hadron momentum (nucleon transverse spin) vector w.r.t.~the lepton scattering plane.  We generated EIC pseudo-data for both transversely polarized proton and $^{3\!}He$ beams with charged pions detected in the final state and applied the JAM20 cuts of $0.2 < z< 0.6,\; Q^2>1.63\,{\rm GeV^2}, \;{\rm and}\;0.2 <P_{hT}<0.9 \,{\rm GeV}$.  Table~\ref{t:data} summarizes the data used in our fit, which includes a total of 8223 EIC pseudo-data points on the Collins effect in SIDIS plus the 517 SSA data points in the original JAM20 global analysis. The EIC pseudo-data covers multiple CM energies $\sqrt{S}$ based on the energy of the electron beam $E_e$ and nucleon beam $E_N$:~$\sqrt{S}\approx 2\sqrt{E_eE_N}$. 
The pseudo-data was generated with pythiaeRHIC~\cite{pythiaerhic} that uses {\sc pythia} 6.4~\cite{Sjostrand:2006za} as an event generator. Realistic EIC detector acceptances and momentum smearing were implemented via the eic-smear package~\cite{eicsmear} and is predominantly based on the expected resolutions that are discussed in the EIC handbook~\cite{handbook}. For pion identification, the momentum and rapidity ranges that evolved from the EIC user group Yellow Report effort~\cite{AbdulKhalek:2021gbh} were used. The proton and $^{3\!}He$ polarizations were assumed to be 70\%, and the uncertainties were scaled to accumulated luminosities of 10 fb$^{-1}$ for each beam energy sample. In the case of $^{3\!}He$, it was assumed that the two protons can be tagged in the very forward instrumentation, and was thus simulated by generating $e+n^\uparrow$ data after taking into account the neutron polarization in $^{3\!}He$. The uncertainties on the expected SSAs were evaluated by re-weighting the unpolarized simulations based on the phenomenological results of Ref.~\cite{Anselmino:2008sga} and extracting the reconstructed asymmetries. As a crude measure of detector smearing and acceptance effects in a real detector, the differences between extracted asymmetries using perfectly tracked and smeared values were assigned as systematic uncertainties. This tries to conservatively mimic the uncertainties that may be related to the unfolding of smearing and particle mis-identification in an actual detector.
\begin{table*}[t]
\centering
\begin{tabular}{ |c|c|c|c| }

\hline
\multicolumn{4}{|c|}{\bf EIC Pseudo-data}
\\\hline \hline

{\bf Observable} & 
{\bf Reactions} & 
{\bf CM Energy $\boldsymbol{\big(\!\!\sqrt{S}\big )}$}&
$\boldsymbol{N_{\rm pts.}}$
\\ \hline
\multirow{14}{*}{Collins (SIDIS)}

&
\multirow{8}{*}{$e+p^\uparrow\to e+ \pi^\pm +X$}  &
\multirow{2}{*}{$141\,{\rm GeV}$} & $756\;(\pi^+)$ \\ 
&   &   & $744\;(\pi^-)$\\ \cline{3-4}

&  &   \multirow{2}{*}{$63\,{\rm GeV}$} & $634\;(\pi^+)$
\\
&   &   & $619\;(\pi^-)$\\ \cline{3-4}

&  &   \multirow{2}{*}{$45\,{\rm GeV}$} & $537\;(\pi^+)$
\\
&   &   & $556\;(\pi^-)$\\ \cline{3-4}

&  &   \multirow{2}{*}{$29\,{\rm GeV}$} & $464\;(\pi^+)$
\\
&   &   & $453\;(\pi^-)$\\ \cline{2-4}

&
\multirow{6}{*}{$e+^{3\!\!\!}He^\uparrow\to e+ \pi^\pm +X$}  & 
\multirow{2}{*}{$85\,{\rm GeV}$} & $647\;(\pi^+)$
\\ 
&   &   & $650\;(\pi^-)$\\ \cline{3-4}

&  &   \multirow{2}{*}{$63\,{\rm GeV}$} & $622\;(\pi^+)$
\\
&   &   & $621\;(\pi^-)$\\ \cline{3-4}

&  &   \multirow{2}{*}{$29\,{\rm GeV}$} & $461\;(\pi^+)$
\\
&   &   & $459\;(\pi^-)$\\ \hline

\multicolumn{1}{c}{} & \multicolumn{1}{c}{} & \multicolumn{1}{|r|}{\bf Total EIC $\boldsymbol{N_{\rm pts.}}$} & {\bf 8223} \\ \cline{3-4}

\multicolumn{4}{c}{}\\
\multicolumn{4}{c}{}\\
 
\hline
\multicolumn{4}{|c|}{\bf JAM20 \cite{Cammarota:2020qcw}}
\\\hline \hline

{\bf Observable} & 
{\bf Reactions} & 
{\bf Experimental Refs.} &
$\boldsymbol{N_{\rm pts.}}$
\\\hline

Sivers (SIDIS)  &
$e+(p,d)^\uparrow\to e+\pi^\pm/\pi^0+X$  & 
\cite{Airapetian:2009ae, Alekseev:2008aa,Adolph:2014zba}& $126$ 
\\

Sivers (DY)  &
$\pi^-\!+p^\uparrow\to \mu^+ \!+ \mu^-+X$  & 
\cite{Aghasyan:2017jop} & 
$12$ 
\\ 

Sivers (DY)  &
$p^\uparrow +p \to W^\pm/Z+X$  & 
\cite{Adamczyk:2015gyk}& 
$17$  

\\ \hline

Collins (SIDIS)  &
$e+(p,d)^\uparrow\to e+\pi^\pm/\pi^0+X$  & 
\cite{Airapetian:2010ds, Alekseev:2008aa,Adolph:2014zba}& $126$
\\

Collins (SIA)  &
$e^++e^-\to \pi^+ \!+ \pi^- +X$  & 
\cite{Seidl:2008xc,TheBABAR:2013yha,Aubert:2015hha,Ablikim:2015pta} & 
$176$ 
\\\hline

$A_N$  &
$p^\uparrow + p\to \pi^\pm/\pi^0 + X$  & 
\cite{Lee:2007zzh,Adams:2003fx,Abelev:2008af,Adamczyk:2012xd} & 
$60$ 
\\[0.05cm]
\hline 

\multicolumn{1}{c}{} & \multicolumn{1}{c}{} & \multicolumn{1}{|r|}{\bf Total JAM20 $\boldsymbol{N_{\rm pts.}}$} & {\bf 517} \\ \cline{3-4}

\end{tabular}
\caption{
    Summary of the data used in our analysis, including the number of points ($N_{\rm pts.}$) in each reaction.   (Top) EIC pseudo-data for the Collins effect in SIDIS for different polarized beam types, CM energies, and final states. (Bottom) Data used in the original JAM20 global analysis of SSAs.} 
\label{t:data}
\end{table*}

\section{EIC Phenomenological Results \label{s:pheno}}
We begin by briefly discussing the methodology of the JAM20 global analysis, which serves as the baseline for our impact study, and refer the reader to Ref.~\cite{Cammarota:2020qcw} for more details.  We employ a Gaussian parametrization for the transverse
momentum dependence of the TMD PDFs and FFs.  In particular, for the transversity TMD PDF we have
\begin{eqnarray}
h_1^q(x,k_T^2) 
&\!\!\!=\!\!\!& h_1^q(x)\ \frac{1}{\pi\avkT_{h_1}^q}\;
{\exp\left[{-\frac{\kT^2}{\avkT_{h_1}^q}}\right]}\,,
\label{eq:f1h1-gauss}
\end{eqnarray}
with $q$ being a quark flavor, and $\avkT_{h_1}^q$ the transverse momentum width.  Note that $\vec{k}_T$ is the transverse momentum of the struck quark. For $h_1^q(x)$ we only
allow $q=u,d$ and explicitly set antiquark functions to zero.  Even though an important goal of the EIC will be to constrain the sea quark transversity PDFs, for the tensor charge their inclusion is expected to have a small effect.  Lattice QCD finds that contributions from disconnected diagrams to the tensor charge are about two orders of magnitude smaller than connected diagrams~\cite{Gupta:2018qil,Alexandrou:2019brg}. 
\begin{figure}[t!]
\centering
\includegraphics[width=0.85\textwidth]{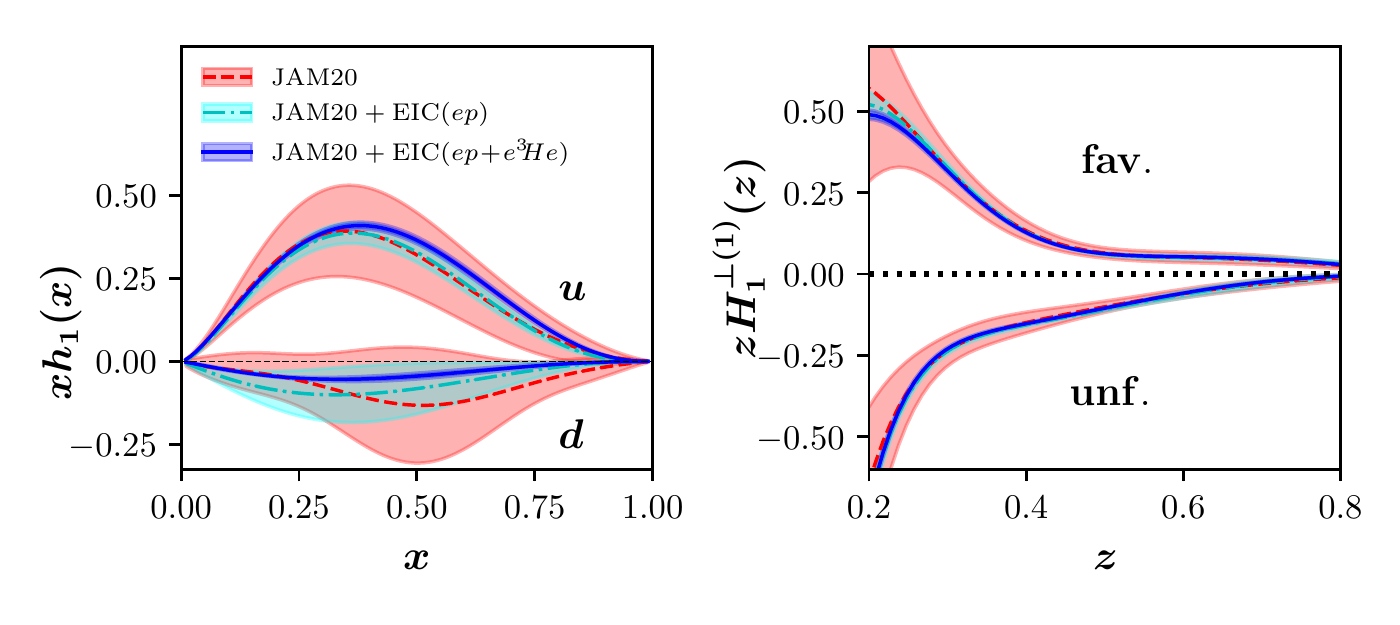}
\includegraphics[width=0.5\textwidth]{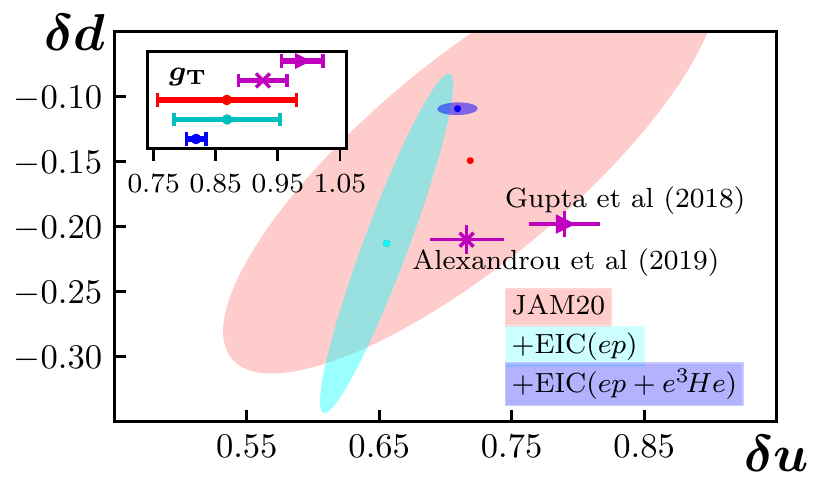}
\caption{(Top) Plot of the transversity function for up and down quarks as well as the favored and unfavored Collins function first moment from the JAM20 global analysis~\cite{Cammarota:2020qcw} (light red band with the dashed red line for the central value) as well as a re-fit that includes EIC Collins effect pion production pseudo-data for a proton beam only (cyan band with the  dot-dashed cyan line for the central value) and for both proton and $^{3\!}He$ beams together (blue band with the solid blue line for the central value). (Bottom) Individual flavor tensor charges $\delta u$, $
\delta d$ as well as the isovector charge $g_T$ for the same scenarios.  Also shown are the results from two recent lattice QCD calculations~\cite{Gupta:2018qil,Alexandrou:2019brg} (purple).  All results are at $Q^2=4\,{\rm GeV^2}$ with error bands at $1$-$\sigma$ CL.}
\label{f:qcf_gT_EIC}
\end{figure}
In addition, if one assumes a symmetric sea, then antiquark contributions cancel when calculating $g_T$, as one can see from Eq.~(\ref{e:gT}).
The Collins TMD FF is parametrized as
\begin{eqnarray}
H_1^{\perp h/q}(z,p_T^2)
&\!\!\!=\!\!\!& \frac{2 z^2 M_h^2}{\pi\left(\avpperp{}^{h/q}_{H_1^\perp}\right)^2}\,
    H_{1\, h/q}^{\perp (1)}(z)\
{\exp\left[{-\frac{\pperp^2}{\avpperp{}^{h/q}_{H_1^\perp}}}\right]}\,,
\label{e:collins}
\end{eqnarray}
where $z$ is the hadron momentum fraction, $\avpperp{}^{h/q}_{H_1^\perp}$ is the transverse momentum width, and $M_h$ is the produced hadron mass. Note that $\vec{p}_T$  is the transverse momentum of the produced hadron with respect to the fragmenting parton.
We allow for favored  and unfavored Collins functions.

The Gaussian transverse momentum parametrizations (\ref{eq:f1h1-gauss}), (\ref{e:collins}) of JAM20 do not have the complete features of
TMD evolution~\cite{Sun:2013dya,Kang:2014zza,Kang:2015msa,Echevarria:2014xaa,Kang:2017btw} and instead assume most of the transverse momentum is non-perturbative and thus related to intrinsic properties of the colliding hadrons rather than to hard gluon radiation.
The JAM20 analysis also implemented a DGLAP-type evolution for the collinear twist-3  functions analogous to Ref.~\cite{Duke:1983gd}, where a
double-logarithmic $Q^2$-dependent term is explicitly added to the
parameters.  Such collinear twist-3 functions arise from the operator product expansion (OPE) of certain transverse-spin dependent TMDs (e.g., $H_{1}^{\perp (1)}(z)$ enters the OPE of the Collins TMD FF~\cite{Kang:2015msa}).  For the collinear twist-2 PDFs and FFs (e.g., $f_1(x)$, $h_1(x)$, and $D_1(z)$),  the standard leading order DGLAP evolution was used.  The fact that current data on SSAs can be described with a simple Gaussian ansatz highlights the need for the tremendous $Q^2$ lever arm of the EIC.  The ability to span several decades in $Q^2$ will help constrain the exact nature of TMD evolution and study the interplay between TMD and collinear approaches.

Our study was conducted using replicas from the JAM20 analysis as priors in a fit of all the data in Table~\ref{t:data} (8740 total points).  The results for the impact on the up and down transversity PDF $h_1(x)$ as well as the Collins function first moment $H_1^{\perp(1)}(z)$ are shown in the top panel of Fig.~\ref{f:qcf_gT_EIC}.  One clearly sees a drastic reduction in the transversity uncertainty band once EIC data is included compared to the original JAM20 results. Even the uncertainties for the Collins function decrease noticeably in the smaller-$z$ region. This will allow for a more stringent test of the universality of the Collins function between SIDIS, electron-positron annihilation, and proton-proton collisions~\cite{Metz:2002iz, Collins:2004nx, Gamberg:2008yt, Yuan:2007nd, Meissner:2008yf,Gamberg:2010uw,Boer:2010ya}.  We emphasize that the $^{3\!}He$ data is crucial for a precise determination of the down quark transversity PDF and for up and down flavor separation, enabling a higher decorrelation between $\delta u$ and $\delta d$.  Specifically, the Pearson correlation coefficients were found to be
$$
\rho[\delta u, \delta d]\equiv \frac{\langle \delta u\cdot\delta d\rangle-\langle\delta u\rangle\langle\delta d\rangle}{\Delta(\delta u)\Delta(\delta d)} = 
	\begin{cases}
	\,0.80 & {\!\!\!\rm for\; JAM20}\,, \\
	\, 0.93 & {\!\!\!\rm for\; JAM20+EIC}(ep)\,,\\
	\, 0.043 & {\!\!\!\rm for\; JAM20+EIC}(ep\!+\!e\hspace{0.025cm}^{3\!}He)\,, 
	\end{cases}
$$
where $\langle \cdots \rangle$ is the average value over all replicas, and $\Delta(\cdots)$ is the uncertainty (standard deviation) of the calculated tensor charge.  (The  correlation coefficient $\rho$ can be in the range $[-1,1]$, where $\rho=\pm1$ indicates $100\%$ correlation (anti-correlation) and $\rho=0$ indicates zero correlation.)

Moreover, the well-constrained up and down $h_1(x)$ translate into very precise calculations of $\delta u$, $\delta d$, and $g_T$, as shown in the bottom panel of Fig.~\ref{f:qcf_gT_EIC}.  We find all relative errors are now  $\lesssim 5\%$: $\delta u = 0.709(15), \delta d  = -0.109(5), g_T= 0.818(16)$.  One can see the increase in precision of the extracted $\delta u$ due to the proton EIC data and further dramatic reduction of errors, in particular for $\delta d$ and $g_T$, in a combined analysis of proton and $^{3\!}He$ EIC data.  From the two lattice QCD calculations at the physical point~\cite{Gupta:2018qil,Alexandrou:2019brg} that are also included in that plot, we can conclude that EIC data will allow for phenomenological extractions of the tensor charges to be as precise as current lattice results.  Thus, the EIC will provide a unique opportunity to explore the possible tension between these two approaches discussed in Ref.~\cite{Radici:2018iag}.

We have also explicitly shown how the central values for $h_1(x)$, $\delta u$, $\delta d$, and $g_T$ shift as new pseudo-data are included in the fit.  Experimental measurements are related to the extracted functions in a very non-linear manner. The inverse problem of extracting parton distribution and fragmentation functions from experimental data can therefore have multiple solutions. In fact, such a shift is expected when a measurement is performed with a very high precision in a limited kinematical region. The measurement will better constrain  parameters describing this particular kinematical region and will, potentially, distort the extracted functions compared to the baseline functions. Thus, a very precise measurement cannot always guarantee a very accurate extraction of the 
distributions, and multiple experiments, such as EIC and SoLID in this case, should be performed in a wide kinematical region in order to minimize bias and expose any potential tension between data sets. This point will be discussed in more detail later in Sec.~\ref{s:SoLID}.

In order to better understand which kinematical regions for the Collins asymmetry are most important to reduce the uncertainties in the extraction of the transversity and Collins functions, we calculate the ratio of the uncertainty in the JAM20 calculation of $A_{UT}^{\sin(\phi_h+\phi_S)}$ to that of the EIC pseudo-data. 
\begin{figure}[t]
\centering
\includegraphics[width=0.775\textwidth]{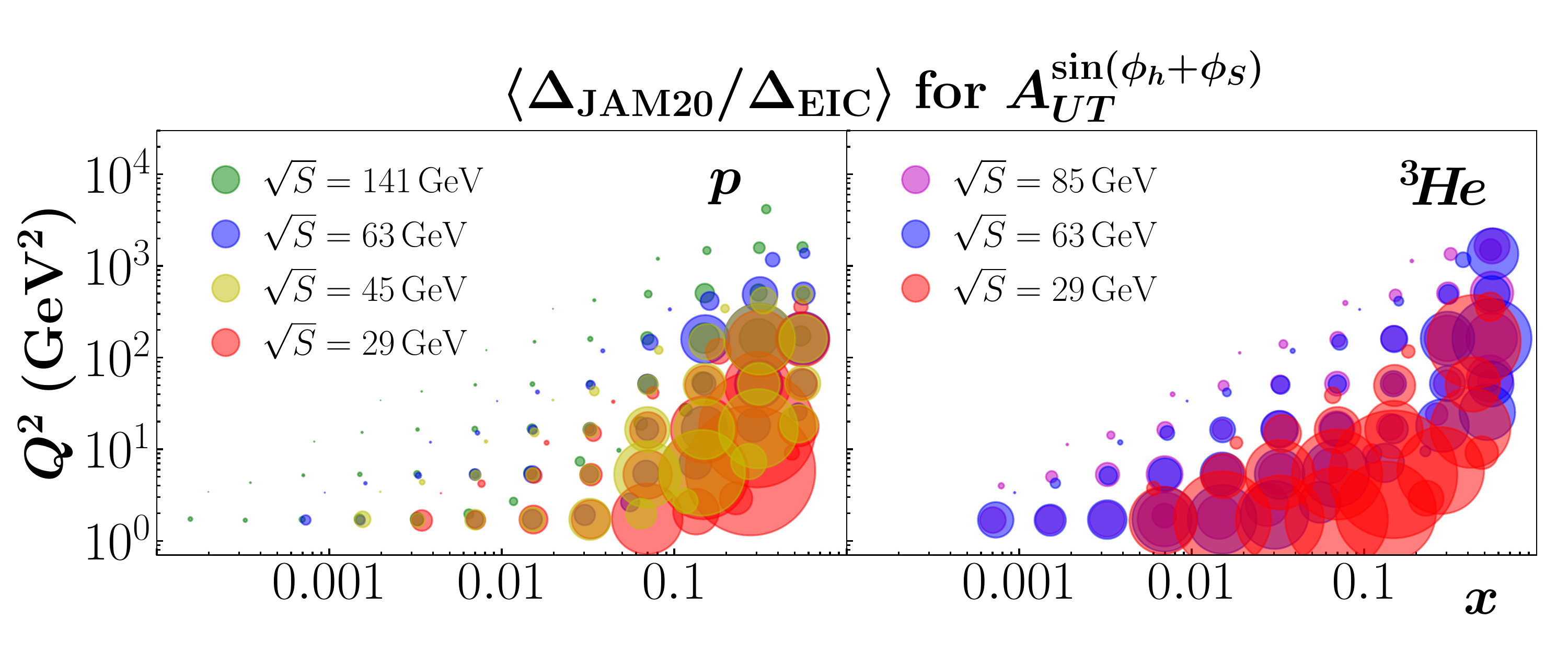}
\caption{The average error ratio as defined in Eq.~(\ref{e:sc_def}) for both proton (left) and $^{3\!}He$ (right) beams for various CM energies. The size of the symbols is proportional to the value of this error ratio.}
\label{f:sens_coeff}
\end{figure}
Since the EIC errors need to be smaller than those from JAM20 in order to obtain more precisely extracted functions, the larger this error ratio, the larger the impact of the new data set on the observable.
We note that $A_{UT}^{\sin(\phi_h+\phi_S)}$ is a function of $(x,Q^2,z,P_{hT})$. Therefore, we define the following average error ratio for each $(x,Q^2)$ bin:
\begin{equation}
\left\langle \Delta_{\rm JAM20}/\Delta_{\rm EIC} \right\rangle \equiv \frac{1}{N_{\rm bin}}\sum_i \left(\Delta_{\rm JAM20}/\Delta_{\rm EIC}\right)_i,
\label{e:sc_def}
\end{equation}
where the sum runs over all points $N_{\rm bin}$ in a given $(x,Q^2)$ bin, including all $(z,P_{hT})$ points in that bin for both $\pi^+$ and $\pi^-$ final states.
The results for $\langle \Delta_{\rm JAM20}/\Delta_{\rm EIC} \rangle$ are shown in Fig.~\ref{f:sens_coeff} for various CM energies for both proton and $^{3\!}He$ beams.  We find that for the proton ($^{3\!}He$) beam, the $x\gtrsim 0.03$ ($x\gtrsim 0.001$) region has the greatest impact on the JAM20 analysis of this observable.  One may ask why more impact is not expected at lower $x$ values for the proton beam.  The reason is the $x$ dependence of the PDFs in JAM20 is parametrized as $\sim Nx^a(1-x)^b$, where $N,a,b$ are free parameters.  Since current SIDIS Collins effect data are in the moderate $x$ region ($0.02 \lesssim x \lesssim 0.3$), the $a$ values in JAM20 are pretty well constrained.  Since at small $x$, the PDFs $\sim x^a$, this leads to reduced uncertainties in this regime even though no data is available there.  Such parametrization bias is unavoidable.  The results of Fig.~\ref{f:sens_coeff} should not be interpreted as diminishing the relevance of the high energy configuration of the EIC for measuring the Collins asymmetry, but rather as an indication of what region most affects our current JAM20 extraction.  Certainly new data in the small $x$ region will influence the value of $a$ and change the inferred shapes of the transversity and Collins functions. In addition, the small $x$ data will reveal a potential sea quark transversity that is not included in our analysis.  Both beams also show significant error reduction over several decades of $Q^2$, highlighting the importance of the tremendous $Q^2$ lever arm of the EIC. We clearly see again in Fig.~\ref{f:sens_coeff} the definite need for the $^{3\!}He$ program at the EIC down to small values of $x$. 
\begin{figure}[htb!]
\centering
\includegraphics[width=0.8\textwidth]{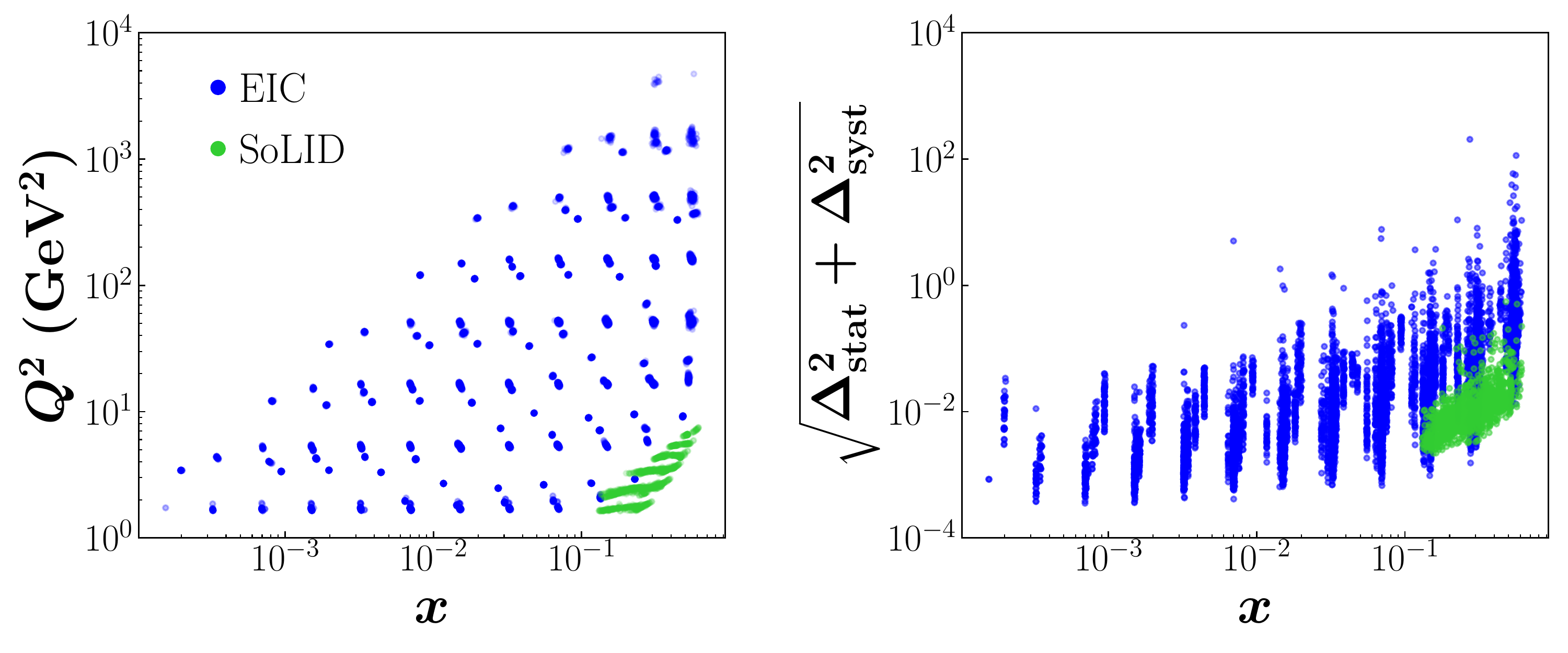}\\
\caption{(Left) Scatter plot of the $x$ and $Q^2$ coverage of the EIC  pseudo-data (blue points) and SoLID pseudo-data (green points). (Right) The quadrature of statistical ($\Delta_{\rm stat}$) and systematic ($\Delta_{\rm syst}$) errors of the pseudo-data for $A_{UT}^{\sin(\phi_h+\phi_S)}$ plotted versus $x$.}
\label{f:EIC_and_SOLID}
\end{figure}
\section{Complementarity of the SoLID Experiment to the EIC \label{s:SoLID}}
\begin{figure}[htb]
\centering
\includegraphics[width=0.6\textwidth]{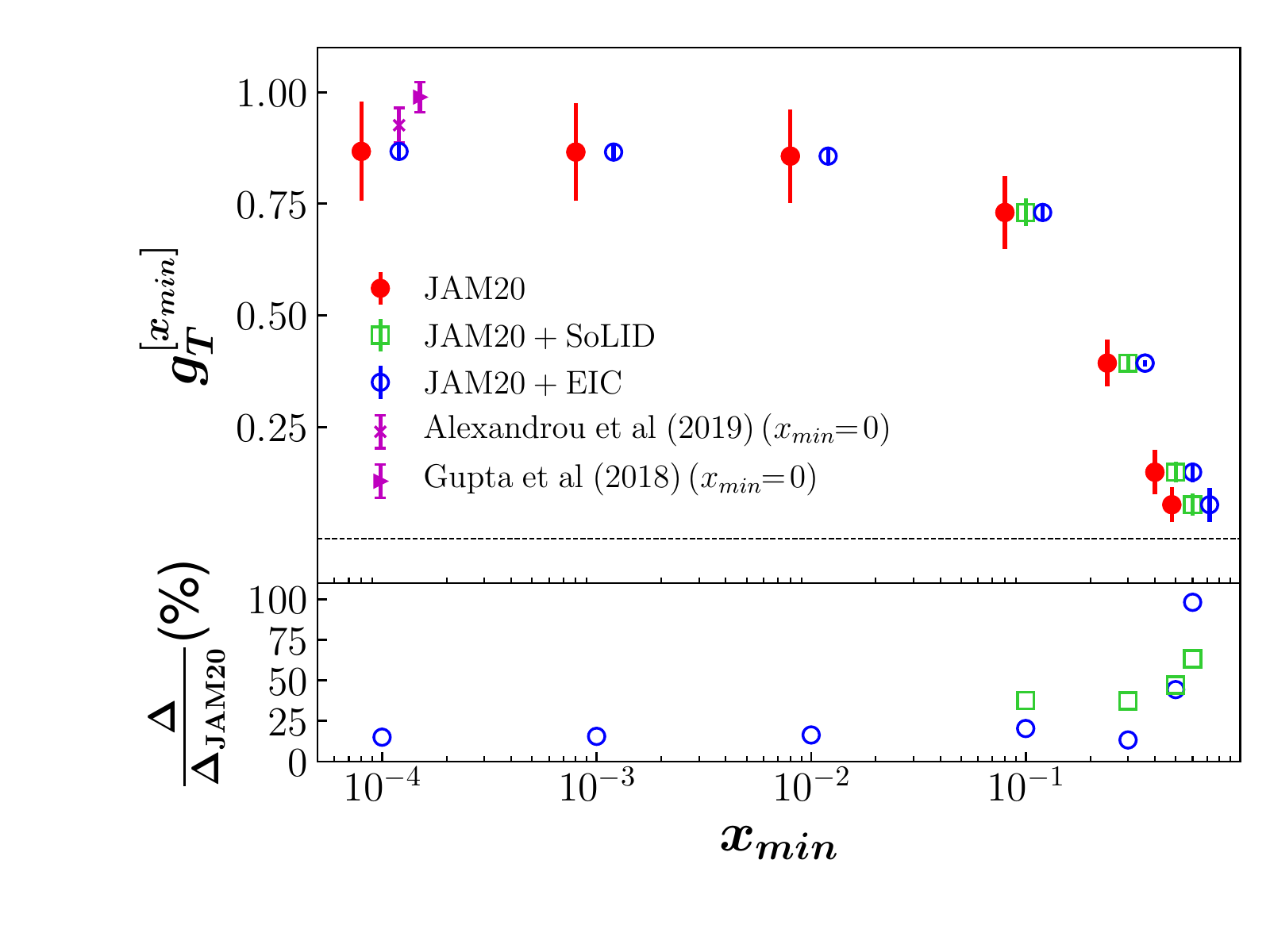}
\caption{ Plot of the truncated integral (as defined in Eq.~(\ref{e:gTxmin})) $g_T^{[x_{min}]}$ vs.~$x_{min}$ for the JAM20 global analysis~\cite{Cammarota:2020qcw} (red points) as well as a re-fit that includes Collins effect pion production pseudo-data (proton and $^{3\!}He$ together) with $x>x_{min}$ from SoLID (green points) and from the EIC (blue points).  The plot also contains two recent lattice QCD calculations~\cite{Gupta:2018qil,Alexandrou:2019brg}.  Note that these lattice data points are for the full $g_T$ integral (i.e., $x_{min}=0$) and have been offset for clarity. Also shown is the ratio $\Delta /\Delta_{\rm JAM20}$ of the uncertainty in $g_T^{[x_{min}]}$ for the re-fit that includes pseudo-data from SoLID (green squares) and for the one that includes pseudo-data from the EIC (blue circles) to that of the original JAM20 fit~\cite{Cammarota:2020qcw}. That is, $\Delta$ in the numerator is either $\Delta_{\rm JAM20+SoLID}$ (for the case of the green squares) or  $\Delta_{\rm JAM20+EIC}$ (for the case of the blue circles).   All results are at $Q^2=4\,{\rm GeV^2}$ with error bars at $1$-$\sigma$ CL.}
\label{f:gTxmin_EIC_SoLID}
\end{figure}
In this section we analyze the impact of pseudo-data from the proposed SoLID experiment at JLab~\cite{Gao:privcom,Chen:2014psa}, compare the results to the EIC case, and discuss the complementary features of these measurements to the EIC. 
SoLID will cover a region of $0.05 \lesssim x \lesssim 0.65$ and $1 \lesssim Q^2 \lesssim 8\,{\rm GeV^2}$.  After the JAM20 data cuts, our study included 526 points for $e+p^\uparrow\to e+ \pi^\pm +X$ (311 for $\pi^+$ and 215 for $\pi^-$) and 696~points for $e+^{3\!\!}He^\uparrow\to e+ \pi^\pm +X$ (412 for $\pi^+$ and 284 for $\pi^-$). The SoLID experiment will use both $8.8 \,{\rm GeV}$ and $11\,{\rm GeV}$ electron beams (CM energy of $\sqrt{S}=4.17\,{\rm GeV}$ and $\sqrt{S}=4.64\,{\rm GeV}$, respectively) for both proton (${\rm NH}_3$) and $^{3\!}He$ targets. Using the tentatively approved running times for both energies and targets, the accumulated luminosities will far exceed the EIC luminosities~\cite{Gao:privcom}. Both EIC and SoLID will be systematics limited in most of their covered kinematical ranges. In the left panel of Fig.~\ref{f:EIC_and_SOLID}, we show the coverage in $x$ and $Q^2$ of the EIC and SoLID pseudo-data. One can see that both facilities cover complementary kinematical regions, i.e., the region of large $x$ and relatively low $Q^2$ for SoLID, and a wider region of $x$, reaching the low values associated with sea quarks and gluons, and large values of $Q^2$ for the EIC. Therefore, for the large-$x$ region the data obtained by SoLID will be important for the detailed exploration of the non-perturbative nature of TMD functions.  The EIC will contribute a substantial $Q^2$ range in the same kinematical domain, which will allow one to study the effects of QCD evolution of TMD functions as well as to constrain them in a wider $x$ range. In addition, these studies will be important for understanding the influence of higher twist corrections, target and produced hadron mass corrections, and the applicability region of TMD factorization.  We also plot in the right panel of Fig.~\ref{f:EIC_and_SOLID} the quadrature of the expected statistical and systematic errors of the EIC and SoLID pseudo-data for $A_{UT}^{\sin(\phi_h+\phi_S)}$. One can see that on average the SoLID pseudo-data will be more precise at larger $x$ due to its higher luminosity.

In Fig.~\ref{f:gTxmin_EIC_SoLID}, we present $g_T^{[x_{min}]}$ vs.~$x_{min}$, where $g_T^{[x_{min}]}$ is the following truncated integral:
\begin{equation}
    g_T^{[x_{min}]} \equiv \int_{x_{min}}^1 \! dx \left[(h_1^u(x)-h_1^{\bar{u}}(x))- (h_1^d(x)-h_1^{\bar{d}}(x))\right]. \label{e:gTxmin}
\end{equation}
We want to study the impact on this quantity only from new data in the region $x>x_{min}$ and eliminate the influence from data with $x<x_{min}$, which could cause an artificial decrease in uncertainties outside the measured region (cf.~the discussion about parametrization bias in connection to Fig.~\ref{f:sens_coeff}).  Therefore, $g_T^{[x_{min}]}$ for JAM20+EIC and JAM20+SoLID is calculated from fits that only include pseudo-data with $x>x_{min}$. We see that the error ratio $\Delta/\Delta_{\rm JAM20}$ increases significantly as one moves towards the edge of the measured region of $x$  ($\sim 0.5-0.6$). As seen in Figs.~\ref{f:sens_coeff}, \ref{f:EIC_and_SOLID}, the EIC still provides coverage around $x\sim 0.5$ with reduced errors from the current JAM20 analysis that at low $Q^2$ are similar to SoLID. Consequently, the EIC is competitive with SoLID for constraining the contribution to $g_T$ from this region.  However, at the very edge of the $x$ phase space ($x\sim 0.6$), where the applicability of the QCD factorization implemented in this letter is yet to be explored, SoLID maintains a reduction in the errors compared to JAM20, whereas the EIC shows no improvement. From Fig.~\ref{f:gTxmin_EIC_SoLID}, we also see that the current JAM20 result only constrains the tensor charge down to $x\sim 0.1$, which accounts for about $75\%$ of the total $g_T$.  Thus, one clearly needs the small-$x$ data at the EIC to fully and precisely determine $g_T$, as Fig.~\ref{f:gTxmin_EIC_SoLID}  highlights.
One also notices that $g_T^{[x_{min}]}$ begins to saturate around $x\sim 0.01$, suggesting that very little tensor charge exists at small $x$.  This observation is consistent with the calculation in Ref.~\cite{Kovchegov:2018zeq} of the small-$x$ asymptotic behavior of the valence transversity TMD PDF.  However, we note that EIC data in the low $x$ region will be needed for the study of the sea quark transversity functions. 

\begin{figure}[t!]
\centering
\includegraphics[width=0.8\textwidth]{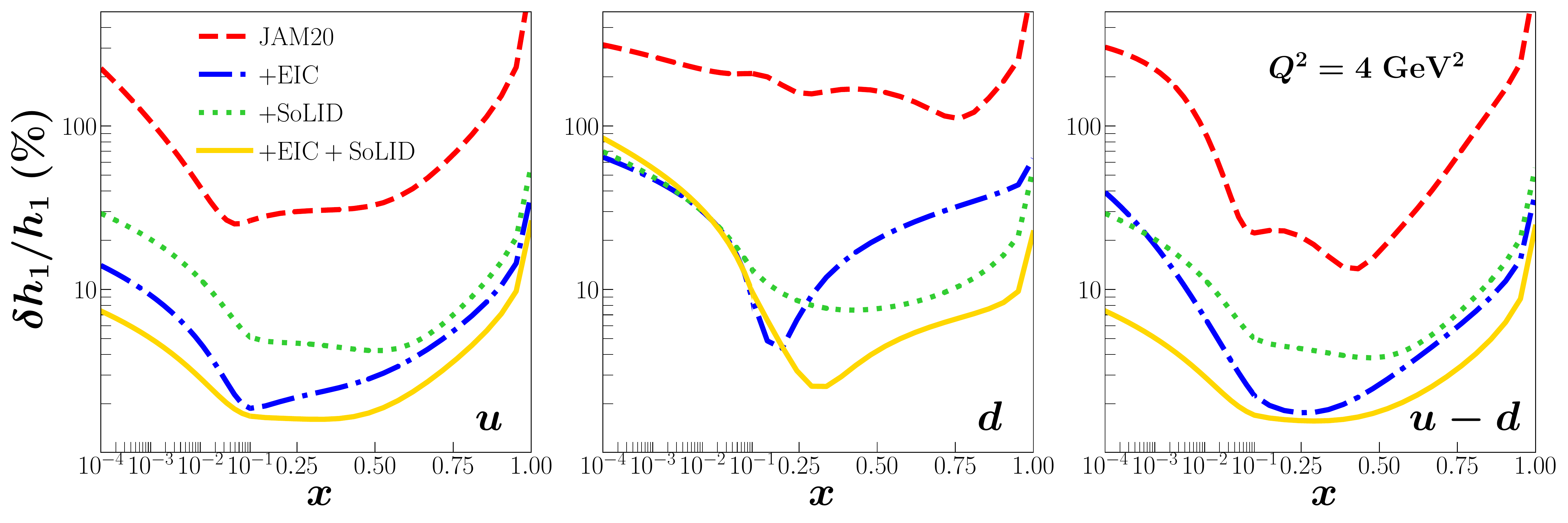}\\
\includegraphics[width=0.65\textwidth]{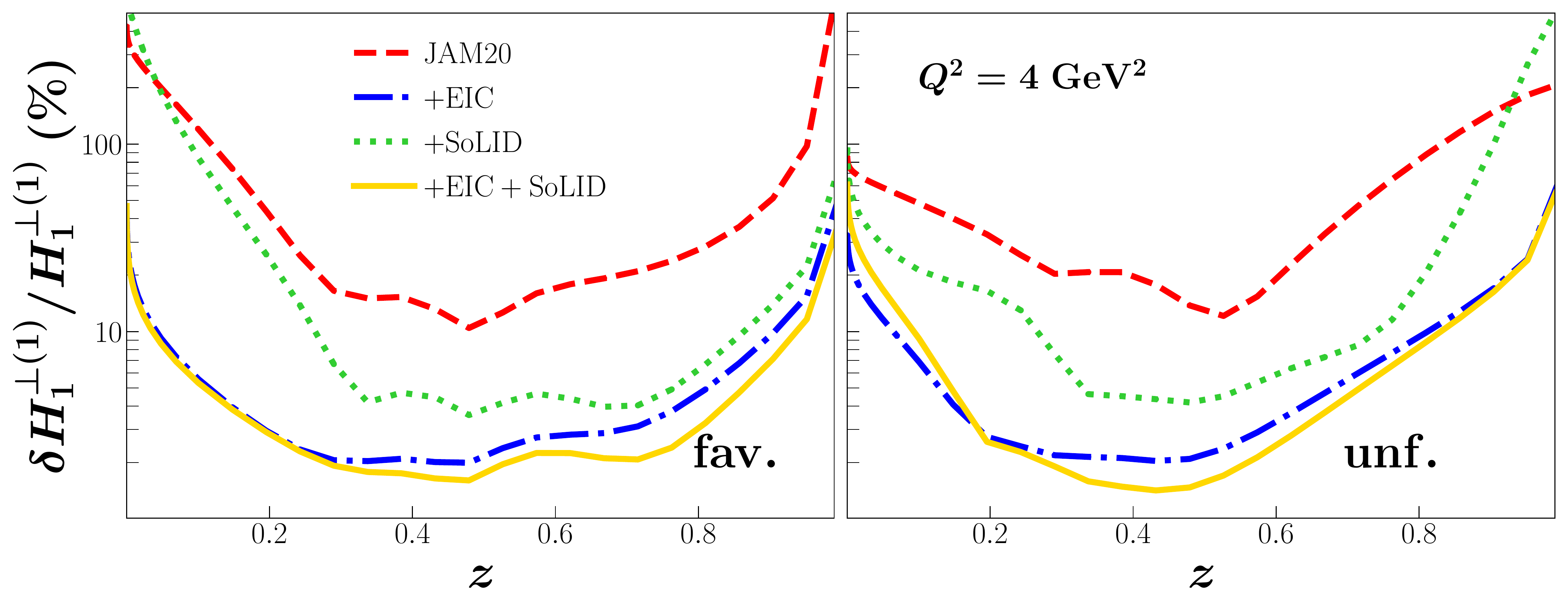}
\caption{(Top) The ratio of the error of transversity to its central value for $u$, $d$, and $u-d$ as a function of $x$ at $Q^2=4$ GeV$^2$ for JAM20 (red dashed line), JAM20+EIC pseudo-data (blue dash-dotted line), JAM20+SoLID pseudo-data (green dotted line), and JAM20+EIC+SoLID pseudo-data (gold solid line). (Bottom) The ratio of the error of the first moment of the Collins FF to its central value as a function of $z$ for favored and unfavored Collins FF.}
\label{f:functions_EIC_SoLID}
\end{figure}

To further compare the EIC and SoLID results, as well as the combined impact from both experiments, in Fig.~\ref{f:functions_EIC_SoLID} we display the relative errors of the transversity function and the Collins function first moment.  For $h_1^u(x)$, we see at larger $x$ the EIC provides a similar reduction in the relative uncertainty as SoLID, and at smaller $x$ the EIC gives a greater decrease.  For $h_1^d(x)$ we find at larger $x$ that SoLID, due to its high luminosity and excellent capabilities with a $^{3\!}He$ target, achieves a greater reduction in the relative uncertainty than the EIC.  Since the size of $h_1^u(x)$ is greater than $h_1^d(x)$, the relative uncertainty for $h_1^{u-d}(x)\equiv h_1^u(x)-h_1^d(x)$ shows a similar behavior as that for $h_1^u(x)$.  The combined fit of including both EIC and SoLID pseudo-data causes a further decrease in the relative uncertainties for transversity in most kinematical regions.  

The Collins FF, since it couples to transversity in the $A_{UT}^{\sin(\phi_h+\phi_S)}$ asymmetry, also experiences a decrease in its relative uncertainties for favored and unfavored fragmentation.  As  previously mentioned, the significant decrease from the EIC for $0.2<z<0.6$ will allow for a check of the universality of the Collins FF between SIDIS, electron-positron annihilation (with forthcoming measurements from Belle-II~\cite{Kou:2018nap}), and proton-proton collisions~\cite{Metz:2002iz, Collins:2004nx, Gamberg:2008yt, Yuan:2007nd, Meissner:2008yf,Gamberg:2010uw,Boer:2010ya}.  SoLID also gives a slight improvement at intermediate $z$ for the Collins function first moment from the one extracted in JAM20, with the sharp rise in the relative error around $z=0.3$ due to the fact that SoLID put a cut of $z>0.3$ on the pseudo-data used for this analysis. The combined analysis of EIC+SoLID is basically identical to the EIC only result.  We note generally in Fig.~\ref{f:functions_EIC_SoLID} that the rapid increase in the relative uncertainties as one moves towards the edges in $x$ or $z$ is indicative of entering an unmeasured region.  The fact that the relative errors are still reduced compared to JAM20 is a consequence of unavoidable parametrization bias, where the impact from regions where new, precise (pseudo-)data are available propagate into kinematics where there is no data.

\begin{figure}[hbt!]
\centering
\includegraphics[width=0.425\textwidth]{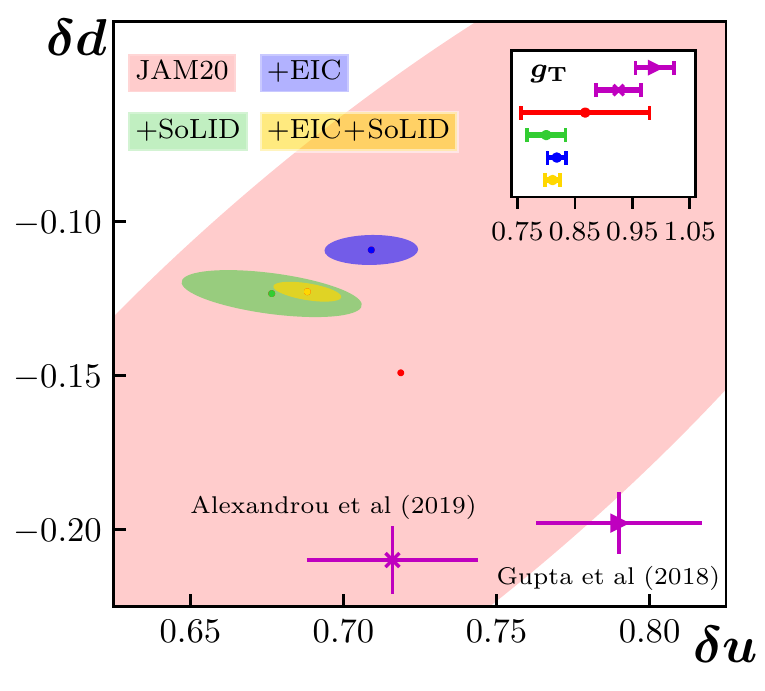}
\caption{Individual flavor tensor charges $\delta u$, $\delta d$ as well as the isovector charge $g_T$ for the same scenarios as Fig.~\ref{f:functions_EIC_SoLID}.}
\label{f:gT_EIC_SoLID}
\end{figure}

In Fig.~\ref{f:gT_EIC_SoLID}, we see a comparison between SoLID and the EIC for $\delta u$, $\delta d$, and the full $g_T$, as well as for the combined fit that included both EIC and SoLID pseudo-data.  We can conclude that SoLID data by itself will also allow for phenomenological extractions of the tensor charges to have similar precision as current lattice results, with relative errors of $\lesssim\! 7\%$:~$\delta u=0.68(3), \delta d = -0.123(8), g_T = 0.80(3)$. The JAM20+EIC+SoLID results give the most precise extractions possible of the tensor charges, more precise than current lattice calculations, with all relative uncertainties now $\lesssim 3\%$:~$\delta u=0.688(11)$, $\delta d=-0.123(3)$, and $g_T=0.811(13)$.

Figure~\ref{f:gT_EIC_SoLID} demonstrates the importance of multiple experimental measurements in a wide kinematical region. The global QCD fits performed on the (pseudo-)data demonstrate that quantities such as tensor charge and the {\em precision} of the extraction depend on many factors: the precision of the data,  the kinematical range of the data, and the flexibility of the model. While the precision of the extraction can be very high, one needs to assure that the {\em accuracy} of the results is also very good. By accuracy we mean the distance from the true value of the measured quantity to the extracted one. One can see from Fig.~\ref{f:gT_EIC_SoLID} that with the generated pseudo-data, our global QCD analysis results in a very precise extraction of the tensor charges for both EIC and SoLID measurements.  However, the 68\% CL regions for the individual flavor charges do not overlap. Thus, the precision of the extracted tensor charges 
may not correspond to the same high accuracy of the result once there are measurements (actual data) from multiple facilities. The reason is an incomplete kinematical region of the experiments and the unavoidable parametrization bias of our extraction. The parametrization bias may be tamed partly by utilizing more flexible parametrizations, such as neural nets. The kinematical coverage of the experiments, on the other hand, is defined by the experimental setup, and it is difficult (if not impossible) to have one experiment cover the whole kinematical region needed for the most accurate extraction. In addition, using data  from only one experiment may bias the extractions, as the systematic errors are quite difficult to account for in an unbiased way. Therefore, multiple experimental measurements covering the largest possible kinematical region are needed to  achieve a precise and simultaneously accurate extraction of the tensor charge. SoLID will offer needed complementary measurements to the EIC in order to test that a consistent picture emerges across multiple experiments on the extracted value of the tensor charge.  Only when a bulk of experiments give consistent central values for quantities of interest, like the tensor charge, can one claim to have accurate results.

\section{Conclusion \label{s:concl}}
In this letter, we have studied the impact on the tensor charge from EIC pseudo-data of the SIDIS Collins effect using the results of the JAM20 global analysis of SSAs~\cite{Cammarota:2020qcw}. Both transversely polarized proton and $^{3\!}He$ beams are considered across multiple CM energies for charged pions in the final state. We find that the EIC will drastically reduce the uncertainty in both the individual flavor tensor charges $\delta u, \delta d$ as well as their isovector combination $g_T$.  The $^{3\!}He$ data is especially crucial for a precise determination of the down quark transversity TMD PDF and for up and down flavor separation.  Consequently, the EIC, from the combined data in  measurements at five different energy settings with transversely polarized proton and $^{3\!}He$ beams, will allow for phenomenological extractions of the tensor charges to be as precise as the current lattice QCD calculations.  This will ultimately show whether a tension exists between experimental and lattice data.  In addition, we performed a similar study on SoLID pseudo-data of the SIDIS Collins effect to be measured in a complementary kinematical region to the EIC and found that the proposed experiment at Jefferson Lab will also significantly decrease the uncertainty in the tensor charge. The combined fit that included both EIC and SoLID pseudo-data provides the best constraint on transversity and the tensor charges, with the results for the latter more precise than current lattice calculations.  We emphasize that a precise measurement cannot always guarantee a very accurate extraction of the 
distributions, and multiple experiments, such as EIC and SoLID, should be performed in a wide kinematical region in order to minimize bias and expose any potential tensions between data sets.  
In order to minimize the bias from the global QCD fit procedure, one may 
ultimately combine the data from different ways of accessing transversity, such as SIDIS single hadron and the di-hadron measurements.
Given that the tensor charge is a fundamental charge of the nucleon and connected to searches for BSM physics~\cite{Courtoy:2015haa,Gao:2017ade,Gonzalez-Alonso:2018omy}, future precision measurements from the EIC and Jefferson Lab sensitive to transversity are of utmost importance and necessary to see if a consistent picture emerges for the value of the  tensor charge of the nucleon.

\section*{Acknowledgments}  
We would like to thank Haiyan Gao, Tianbo Liu, Jian-Ping Chen, and Jianwei Qiu for useful discussions. This work has been supported by the National Science Foundation under Grants No.~PHY-1945471 (Z.K.), No.~PHY-2011763 (D.P.), No.~PHY-2012002 (A.P.), the U.S. Department of Energy, under contracts No.~DE-FG02-07ER41460 (L.G.), No.~DE-AC05-06OR23177 (A.P., N.S.) under which Jefferson Science Associates, LLC, manages and operates Jefferson Lab, and within the framework of the TMD Topical Collaboration. The work of N.S. was supported by the DOE, Office of Science, Office of Nuclear Physics in the Early Career Program.

\bibliographystyle{h-physrev}

\end{document}